\documentclass[aps,prd,twocolumn,showpacs,preprintnumbers,amsmath,superscriptaddress,amssymb,floats,nofootinbib]{revtex4}
\usepackage{graphicx,color}

\newcommand{\black}[1]{{\color{black}{#1} }}
\begin{document}
\newcommand {\T} {{\cal T}}
\newcommand {\DC} {{\delta \chi^2}}
\newcommand {\A} {{A}}
\newcommand {\al} {a}
\newcommand {\bl} {b}
\newcommand {\erf} {\text{erf}}
\newcommand {\ttheta} {\theta} 
\newcommand {\x}  {{\theta_{\text{true}}}}
\newcommand {\y}  {{\theta_{\min}}}
\newcommand {\Tm} {{\Theta_{\min}}}
\newcommand {\Tt} {{\Theta_{\text{true}}}}

\def\Journal#1#2#3#4{{#1} {\bf #2}, #3 (#4)}
\def\NCA{\rm Nuovo Cimento}
\def\NPA{{\rm Nucl. Phys.} A}
\def\NIM{\rm Nucl. Instrum. Methods}
\def\NIMA{{\rm Nucl. Instrum. Methods} A}
\def\NPB{{\rm Nucl. Phys.} B}
\def\PLB{{\rm Phys. Lett.}  B}
\def\PRL{\rm Phys. Rev. Lett.}
\def\PRD{{\rm Phys. Rev.} D}
\def\PRC{{\rm Phys. Rev.} C}
\def\ZPC{{\rm Z. Phys.} C}
\def\JPG{{\rm J. Phys.} G}



\title{Statistical Evaluation of Experimental Determinations of Neutrino Mass Hierarchy}
\date{\today}
\author{X. Qian}\email[Corresponding author: ]{xqian@caltech.edu}
\affiliation{Kellogg Radiation Laboratory, California Institute of Technology, Pasadena, CA}
\author{A. Tan}\email[Corresponding author: ]{aixin-tan@uiowa.edu}
\affiliation{Department of Statistics and Actuarial Science, University of Iowa, Iowa City, IA}
\author{W. Wang}\email[Corresponding author: ]{wswang@wm.edu}
\affiliation{Physics Department, College of William and Mary, Williamsburg, VA}
\author{J. J. Ling}
\affiliation{Brookhaven National Laboratory, Upton, NY}
\author{R. D. McKeown}
\affiliation{Thomas Jefferson National Accelerator Facility, Newport News, VA}
\affiliation{Physics Department, College of William and Mary, Williamsburg, VA}
\author{C. Zhang}
\affiliation{Brookhaven National Laboratory, Upton, NY}

\begin{abstract} 
  Statistical methods of presenting experimental
  results in constraining the neutrino mass hierarchy (MH) are discussed. 
  Two problems are considered and are related to each other: how to report the findings for observed
  experimental data, and how to evaluate the
  ability of a future experiment to determine the neutrino mass
  hierarchy, namely, sensitivity of the experiment. For the first problem where experimental data have
  already been observed,
  the classical statistical analysis involves constructing confidence
  intervals for the parameter $\Delta m^2_{32}$. These intervals are
  deduced from the parent distribution of the estimation of $\Delta
  m^2_{32}$ based on experimental data. Due to existing experimental
  constraints on $|\Delta m^2_{32}|$, the estimation of $\Delta
  m^2_{32}$ is better approximated by a Bernoulli distribution (a
  Binomial distribution with 1 trial) rather than a Gaussian
  distribution. Therefore, the Feldman-Cousins approach needs to be
  used instead of the Gaussian approximation in constructing
  confidence intervals. Furthermore, as a result of the definition of
  confidence intervals, even if it is correctly constructed, its
  confidence level does not directly reflect how much one hypothesis
  of the MH is supported by the data rather than the other
  hypothesis. We thus describe a Bayesian approach that quantifies the
  evidence provided by the observed experimental data through the
  (posterior) probability that either one hypothesis of MH is
  true. This Bayesian presentation of observed experimental results is
  then used to develop several metrics to assess the sensitivity of
  future experiments. Illustrations are made using a simple example with
  a confined parameter space, which approximates the MH determination 
  problem with experimental constraints on the $|\Delta m^2_{32}|$.
\end{abstract}

\maketitle
\thispagestyle{plain}

\section{Introduction}\label{sec:introduction}

Neutrino mass hierarchy (MH), i.e. whether the mass of the third
generation neutrino ($\nu_3$ mass eigenstate) is greater or less than
the masses of the first and the second generation neutrinos ($\nu_1$
and $\nu_2$), is one of the main questions to be answered in the
standard model. Besides its fundamental importance to neutrino
oscillation physics, the resolution of the neutrino MH plays an
important role for the search of neutrinoless double-beta decay, which
would determine whether neutrino is a Dirac or Majorana fermion. With
the recent discovery of a large value of $\sin^{2}2\theta_{13}$ from
Daya Bay~\cite{dayabay,dayabay_cpc,dayabay_proceeding,dayabay_long}, T2K~\cite{t2k}, MINOS~\cite{minos}, Double
Chooz~\cite{doublec}, and RENO~\cite{RENO}, the stage for addressing
the neutrino MH has been set. It became one of the major goals of
current and next generation long baseline neutrino experiments
(T2K~\cite{t2k1}, NO$\nu$A~\cite{nova} and LBNE~\cite{Akiri:2011dv})
and atmospheric neutrino experiments (Super-K~\cite{superk},
MINOS~\cite{minos1}, PINGU~\cite{pingu}, and INO~\cite{ino}).  Meanwhile, the idea of
utilizing a reactor neutrino experiment to determine the MH is also
intensively discussed~\cite{petcov,learned,zhan,xqian,Emilio}.

The objective of this paper is to present appropriate ways to do
statistical analysis that will help determine the neutrino mass
hierarchy.  We start by introducing a few symbols and state the
physics problem in terms of a pair of statistical hypotheses. Let
$m_1$, $m_2$ and $m_3$ denote the masses of the $\nu_1$, $\nu_2$ and
$\nu_3$ mass eigenstate neutrinos, and let $\Delta m_{ij}^2 \equiv
m_i^2-m_j^2$ for $i,j=1,2,3$.  As reviewed in Ref.~\cite{mckeown}, it
is known that $\Delta m_{21}^2>0$ from measurements of solar
neutrinos given the definition of mixing angle $\theta_{12}$. 
Whereas the sign of $\Delta m_{32}^2$ is so far unknown,
and it's common to use NH and IH to denote the two hypotheses, the
normal hierarchy and the inverted hierarchy, respectively:
\begin{equation}\label{eq:hypo}
\begin{cases}
\begin{array}{ll}
  {\rm NH}:& \Delta m_{32}^2>0\;;\\
  {\rm IH}:& \Delta m_{32}^2<0\;.\\
\end{array}\end{cases} 
\end{equation}
A unique feature to the above hypotheses testing problem is that,
there are additional, rather strong information regarding the
parameter $\Delta m_{32}^2$
that need to be taken into account properly.  Actually, based on
previous experiments, a $68\%$ confidence interval
of 
$M^2_{32}\equiv |\Delta m^2_{32}|$ is given by $(2.43\pm0.13)\times
10^{-3}$ eV$^2$~\cite{PDG}.

We will mainly address two aspects of the hypotheses testing
problem. The first one concerns conducting a test after data has been
collected. We discuss a classical testing procedure based on a
$\Delta\chi^{2}$ statistic (Eq.~\ref{eq:chisqmin}), or equivalently,
the procedure of constructing confidence intervals by inverting the
test.
As a matter of fact, the classical procedure is derived upon the
assumption that the best estimator of $\Delta m^2_{32}$ based on
experimental data would follow a distribution that is approximately
Gaussian.  But due to existing constraints on $M^2_{32}$, this
assumption is far from being satisfied. Consequently, actual levels of
the resulting confidence intervals may deviate substantially from
their nominal levels, as we demonstrate in
Sec.~\ref{sec:parameter}. Instead, a general way to construct
confidence intervals that are true to their nominal levels is the
Feldman-Cousins approach~\cite{FC_stat}, which we also illustrate in
detail in Sec.~\ref{sec:parameter}.

Still, there is a fundamental limitation to the use of confidence
intervals. Note that in the MH determination problem, one of the most
crucial questions
is, what is the chance that the MH is indeed NH (or IH) given the
observed experimental data? Classical confidence intervals are not
meant to answer this question directly, whereas credible intervals
reported by a Bayesian procedure is. In Sec.~\ref{sec:bayes}, we
present a Bayesian approach, which effortlessly incorporates prior
information on $M^2_{32}$ and output the easy-to-understand
(posterior) probability of NH and IH to conclude the test.
We will emphasize the importance to differentiate the Bayesian
credible interval from the classical confidence interval.





The second aspect of the hypotheses testing problem that we address
concerns assessment of experiments in their planning stage. It is
critical to evaluate the ``sensitivity'' of a proposed experiment,
i.e., its capability to distinguish NH and IH. Since this evaluation is
performed before data collection, it has to be based on potential data
from the experiment. An existing evaluation method (such as employed in 
\cite{Bernabeu:2010rz,Akiri:2011dv,Blennow:2012gj}) assumes that the
most typical data set under one hypothesis, say NH, happens to have
been observed. Such a data set is referred to as the \textit{Asimov
  data set}~\cite{Cowan}.
The method then calculates $\overline{\Delta\chi^{2}}$, which
  stands for the statistic $\Delta\chi^{2}$ in Eq.~\ref{eq:chisqmin},
  with the extra bar indicating its dependence on the Asimov data
  set. 
It can be seen that
$\overline{\Delta\chi^{2}}$ reflects how much the Asimov data set
under NH disagrees with the alternative model, IH. It is then common
practice to quantify the amount of disagreement by finding the p-value
corresponding to $\overline{\Delta\chi^{2}}$ after comparing it to the
quantiles of a chi-square distribution with one degree of
freedom (choice of MH). Finally, one minus this p-value is sometimes reported as a
quantitative assessment
 of the experiment. 
We will show in Sec.~\ref{sec:parameter} that the comparison of the value of
$\overline{\Delta\chi^{2}}$ to the quantiles of a chi-square
distribution is not justified, 
when previous knowledge impose constraints
on the range of possible values of the parameter $\Delta m^2_{32}$.
         
As an alternative solution, we adopt a Bayesian framework and develop a
set of new metrics for sensitivity to evaluate the potential of
experiments to identify the correct hypothesis.

The paper is organized as follows. In Sec.~\ref{sec:parameter},
we review the steps to construct classical confidence intervals for
the parameter $\Delta m^2_{32}$.
In Sec.~\ref{sec:bayes}, we describe a Bayesian approach that
reports the probability of each hypothesis of MH given observed
data set. We further extend this Bayesian method to help assess the
sensitivity for future experiments. In Sec.~\ref{sec:example}, we
illustrate the Bayesian approach for a simplified version of
the MH problem. In particular, analytical formula of the 
approximations for the probability of the hypotheses, and those 
for the sensitivity metrics are provided.
Also, a numerical comparison is made between the
$\overline{\Delta\chi^{2}}$ based on the Asimov data set and the
sensitivity metrics based on the Bayesian approach. Finally,
discussions and a summary are presented in
Sec.~\ref{sec:discussion} and Sec.~\ref{sec:summary}, respectively.\\

\section{Estimation in constrained versus unconstrained parameter spaces}
\label{sec:parameter}
In this section, we review a classical statistical procedure of
forming confidence intervals.  For the problem of determining the
neutrino mass hierarchy, we demonstrate that the procedure is valid in
one scenario, but fails in another where known constraints on
$M^2_{32}$ are taken into consideration. In the latter case, the
Feldman-Cousins method~\cite{FC_stat} based on Monte Carlo (MC) simulation
is recommended to obtain valid confidence intervals.

Consider a spectrum that consists of $n$ energy bins. Assume that the
expected number of counts in each bin is a function of $\Delta
m^2_{32}$ and a nuisance parameter $\eta$. For simplicity, we denote
$\Delta m^2_{32}$ by $\theta$.  Then for the $i$th bin, let
$\mu_{i}(\theta, \eta)$ and $N_i$ represent the expected and the
observed counts of neutrino induced reactions, respectively.  When
$\mu_{i}$ is large enough, the distribution of $N_i$ can be well
approximated by a Gaussian distribution with mean $\mu_{i}$ and
standard deviation $\sqrt{\mu_{i}}$.
 
Once the data $x=\{N_i, i=1,\ldots,n\}$ are observed, the deviations
from the expected values $\{\mu_i(\theta,\eta), i=1,\ldots,n\}$ are
often calculated to help measure the implausibility of the parameter
$(\theta,\eta)$.
Specifically, when the systematic uncertainties are omitted, and that
certain available knowledge concerning the parameters $\theta$ and
$\eta$ are taken into consideration, 
one useful definition of the deviation is given by
\begin{widetext}
\begin{equation}\label{eq:chi2def}
\begin{array}{lllllll}
\chi^2 (\theta,\eta) &=& \chi^2_{stat}(\theta,\eta)  &+& \chi^2_p (|\theta|) &+& \chi^2_p(\eta)\\
&=& \sum_i \frac{(N_i-\mu_{i}(\theta, \eta))^2}{(\delta N_i)^2} &+& \frac{(|\theta| - |\theta_0|)^2}{(\delta|\theta|)^2} 
&+& \frac{(\eta-\eta_0)^2}{(\delta \eta)^2}\,.  
\end{array}
\end{equation}
\end{widetext}
Here, the general notation $\delta w$
represents 
the standard deviation of a variable $w$.  So $\delta N_i=\sqrt{\mu_{i}}$, and the corresponding $\chi^2_{stat}$ term is called the Pearson's chi-square.
Also, note that $|\theta|=M^2_{32}$, and it is taken from~\cite{PDG}
that $|\theta_0|=2.43\times10^{-3}$ eV$^2$ and $\delta |\theta|
=0.13\times10^{-3}$ eV$^2$.

\begin{table*}[ht!]
\begin{center}
\begin{tabular}{|c|c|c|c|c|c|c|}
\hline
Case & $\Delta\chi^{2}_{min}(\theta_{true})$ & $\theta_{min}$ & distribution & $\Delta \chi_{min}^2 \leq 1$ & $\Delta \chi^2_{min} \leq 4$& $\Delta \chi^2_{min} \leq 9$ \\
     & distribution       & distribution & parameter & confidence & confidence & confidence\\
     &                    &              & within this example & level & level & level \\\hline
I & Chi-square & Gaussian & mean = 1 and $\sigma=0.67$& 68.27\%& 95.48\%& 99.73\% \\\hline
II & -  & Bernoulli & $p=0.0679$ & 95.12\%& 98.48\%& 99.86\%\\
\hline
\end{tabular}
\end{center}
\caption{\label{table:par}Confidence levels for various of $\Delta \chi^2_{min}$ region
  for the Gaussian and the Bernoulli distribution from MC.  In Case I, the mean and the standard 
  deviation of the Gaussian distribution is found to be about 1 and $\sigma=0.67$
  respectively.  In Case II, the parameter $p$ of the Bernoulli distribution (e.g. 
  percentage of $\theta_{min}<0$) is is found to be about 6.8\%. }
\end{table*}

Based on Eq.~\ref{eq:chi2def} and a standard procedure discussed in
Ref.~\cite{PDG}, confidence intervals can be obtained for the
parameter of interest $\theta$ ($\Delta m^2_{32}$), the sign of which
is an indicator of the neutrino MH.
First, define $\theta_{\min}$ to be the best fit to the data in the
sense that $(\theta_{\min},\eta_{\min})=\arg\min_{\theta,\eta}
\chi^2(\theta,\eta)$ where the minimum is taken over $\Theta \times
H$, the space of all possible values of $(\theta,\eta)$. Here, the
general notation $\arg\min_{w} h(w)$ denotes the value of $w$ which
corresponds to the minimum of the given function $h$.  Note that
  $\theta_{\min}$ suggested by the observed data set will not be
  exactly the true value of the parameter $\theta$, and a repetition
  of the experiment would yield a data set that corresponds to a
  different $\theta_{\min}$. So instead of reporting only
  $\theta_{\min}$, it is more rational to report a set of probable
  values of $\theta$ that fit the observed data not too much worse
  than that of the best fit, and state how trust worthy the set
  is. Indeed, for any given $\theta$, let $\eta_{\min}(\theta)=\arg
  \min_{\eta} \chi^2(\theta,\eta)$, and
define
\begin{equation}\label{eq:chisqmin}
\Delta\chi^{2}_{min} (\theta) \equiv \chi^2(\theta,\eta_{min}(\theta))-\chi^2(\theta_{min},\eta_{min}),
\end{equation}
then 
a \textit{level $\al$ confidence interval} based on Eq.~\ref{eq:chisqmin} is 
defined to be
  \begin{equation}\label{eq:CI}C_\al=\{\theta \in \Theta: \Delta\chi^{2}_{min} (\theta) \leq t_\al \}\,,\end{equation}
  where we use the standard set-builder notation $\{h(w):
  \text{restriction $w$}\}$ to denote a set that is made up of all the
  points $h(w)$ such that $w$ satisfies the restriction to the right
  of the colon. The key in constructing Eq.~\ref{eq:CI} is to specify
  the correct threshold value $t_\al$ for a given confidence level
  $\al$. (See the final paragraph of this section for a more detailed description of what confidence level means.)   Most commonly examined confidence levels use $\al= 68.27\%
  (1\sigma)$, $95.45\% (2\sigma)$, $99.73\% (3\sigma)$, which are often
  linked to threshold values $t_\al=$ 1, 4, 9
  respectively~\cite{PDG}.  Note that these three values are the
  $68.27\%$, $95.45\%$ and $99.73\%$ quantiles of the chi-square
  distribution with one degree of freedom, respectively. They are used
  as threshold values because the parameter space $\Theta$ is of
  dimension one and that, under certain regularity conditions,
  $\Delta\chi^{2}_{min} (\theta)$ would follow approximately a
  chi-square distribution with one degree of freedom when $\theta$ is
  the true parameter value. This procedure and its extensions to cases
  where $\theta$ is of higher dimension have been successfully applied
  in many
  studies~\cite{Guo:2007ug,Ardellier:2006mn,Ahn:2010vy,Huber:2004xh,
    Huber:2004ka,Huber:2007ji,Bernabeu:2010rz,Akiri:2011dv,Blennow:2012gj,Schwetz}
  in order to constrain various parameters in the neutrino physics.

Although this procedure has been widely used in analyzing experimental
data, 
note that it is not universally applicable. Its limitations has been addressed by Feldman and
Cousins~\cite{FC_stat}. Below, we illustrate this point through a simple MC simulation study. 
It will be shown that, in 
a situation that is
similar in nature to the MH determination problem in Eq.~\ref{eq:hypo} where there exist special constraints on the possible values of $\theta$, the aforementioned threshold values based on chi-square approximation could result in bad confidence intervals. That is, the actual coverage probabilities of the intervals strongly disagree with their nominal levels. 

\begin{figure*}
\centering
\includegraphics[width=150mm]{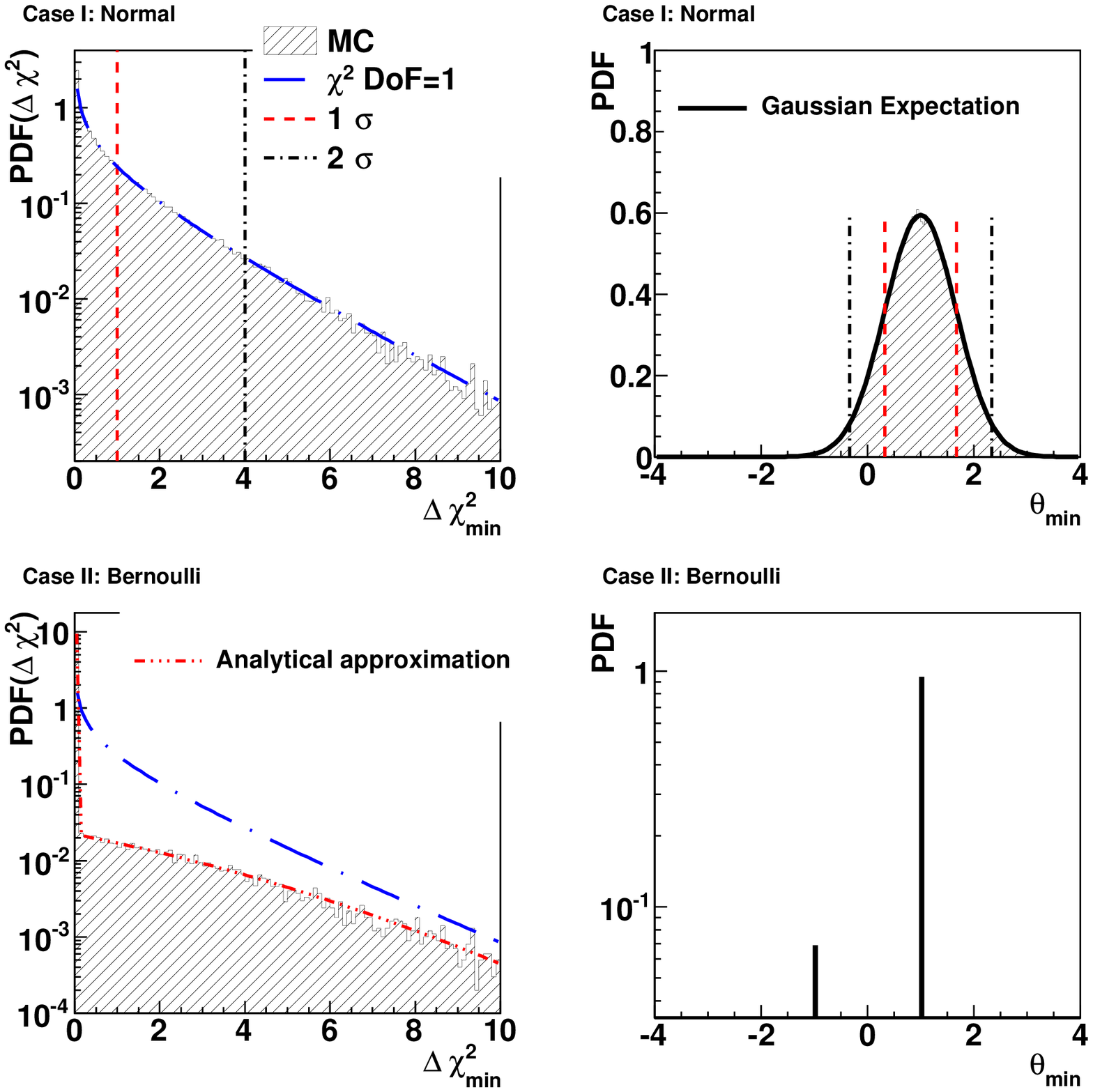}
\caption{(color online) Distributions of $\Delta\chi^{2}_{min}(\theta_{0})$ 
and $\theta_{min}$ for case I and case II with 100,000 MC samples. 
The $\theta_{min}$ distribution of case I (top right) and case II (bottom right)
are a Gaussian and a Bernoulli distribution, respectively. The $\Delta
\chi^{2}_{min}(\theta_{0})$ distribution of case I (top left) is
consistent with the chi-square distribution with degree of freedom
one. The commonly used $1\sigma$, ($68.27\%$ confidence level) and 
$2\sigma$, ($95.45\%$ confidence level) regions are labelled 
with red dashed and black dash-dotted lines for case I. 
The $\Delta \chi^{2}_{min}(\theta_{0})$ distribution of
case II (bottom left) strongly deviates from the chi-square distribution. 
In case II, we also show the analytical approximation (derived 
in Appendix.~\ref{sec:ap1}) of the distribution of $\Delta \chi^2_{min}$. We should 
emphasize while chi-square distribution does not depend on any additional parameter (other than
$\Delta \chi^2_{min}$), the analytical approximation depends on $\overline{\Delta \chi^2}$.}
\label{fig:example}
\end{figure*}
 
In the simulation, we set $n=10$, $\mu_i(\theta) = 1000 + 15\cdot
\theta$ for $i=1,\ldots, n$. (Here, no nuisance parameter $\eta$ is
introduced, and all the expected bin counts are assumed equal for
simplicity. Nevertheless, these assumptions are not essential to the purpose of our simulation.)
The following two cases are investigated: 
\begin{itemize}
\item Case I: $\Theta=(-\infty,\infty)$, 
\item Case II: $\Theta=\{-1, 1\}$.
\end{itemize}
 Case I is a typical situation where nothing was known about
$\theta$ before the current experiment, whereas case II 
is designed to imitate the situation where existing measurements of
  $|\theta|=M^2_{32}$ are very accurate at around $2.43\times10^{-3}$
  eV$^2$, and we simply denoted this value to $1$ for clarity of  presentation. Further, the definition of deviation analogue to
  Eq.~\ref{eq:chi2def} \black{is taken to be} 
  $\chi^2 (\theta)=\sum_i \frac{(N_i-\mu_{i}(\theta))^2}{\mu_{i}(\theta)}$ 
  for case I. 
  For case II, the chi-square definition is $\chi^2 (\theta)=\sum_i \frac{(N_i-\mu_{i}(\theta))^2}{\mu_{i}(\theta)}+\frac{(|\theta| - |\theta_0|)^2}{(\delta|\theta|)^2}$ with experimental constrains on $|\theta|$. It is then reduced to
 $\chi^2 (\theta)=\sum_i \frac{(N_i-\mu_{i}(\theta))^2}{\mu_{i}(\theta)}$ with
$\theta$ being only 1 or -1.


Under each case, we set the true value of $\theta$ to be $\theta_0=1$,
based on which 100,000 MC samples are simulated, denoted by
$\{N_1^{(j)},\cdots,N_{10}^{(j)}\}$ for $j=1,\ldots, 100000$. Then for
the $j$th sample, confidence intervals of levels $\al= 68.27\%$,
$95.45\%$, $99.73\%$ are constructed according to Eq.~\ref{eq:CI}
using threshold values 1, 4, 9, respectively. Finally, at each of the
three levels, we record the proportion of confidence intervals out of
the 100,000 that include the truth $\theta_0=1$. The results are
reported in the last three columns of Table~\ref{table:par}. It can be
seen that, in case I, the actual coverage probabilities closely match
the nominal levels. However, in case II, the actual coverage
probabilities are always higher!

Without too much technical detail, we try to explain the reason why
the chi-square procedure produced valid confidence intervals for case
I, but not for case II. In general, having observed data $x$ from a
parametric model $P(x|\theta)$, a sensible test for a pair of 
hypotheses, $H_0: \theta \in\Theta_0$ and $H_1:\theta \in
\Theta-\Theta_0$ (the counterpart of
$H_0$), 
is the likelihood ratio test that is based on the test statistic
\begin{equation}\label{eq:dchi22}
  \Delta \chi^2_{min} \equiv -2 {\rm log} \left( \frac{P(x|\theta_{0,\min})}{P(x|\theta_{\min})}\right),
\end{equation} 
where $\theta_{0,\min}=\arg\min_{\{\theta \in \Theta_0\}}
P(x|\theta)$, and $\theta_{\min}=\arg\min_{\{\theta \in \Theta\}}
P(x|\theta)$ are the best fit over the null parameter set $\Theta_0$
and the full parameter set $\Theta$, respectively. If the observed data
$x$ yields a large $\Delta \chi^2_{min}$, it means that $\Theta_0$ is
implausible, which further leads to the rejection of $H_0$.  Note
that, the statistic $\Delta\chi^{2}_{min} (\theta)$ in
Eq.~\ref{eq:chisqmin} is a special case of Eq.~\ref{eq:dchi22} with
$\Theta_0$ consisting of a single point, $\theta$.
  
In order to determine the correct threshold values in rejecting, or
equivalently, in constructing confidence intervals defined by
Eq.~\ref{eq:CI}, the distribution/quantiles of $\Delta\chi^{2}_{min}
(\theta_0)$ considering all possible data set
needs to be known, when the true parameter value is some
$\theta_0\in \Theta_0$. An important result in statistics, Wilks
Theorem~\cite{Wilks-1938,stat4} states that, under certain regularity
conditions, $\Delta\chi^{2}_{min} (\theta_0)$ follows approximately a
chi-square distribution with degree of freedom equal to the difference
between the dimension of $\Theta$ and that of $\Theta_0$, when the
data size is large. (In our problem, the data size is simply $\sum_i
N_i$.)  The main regularity conditions are, as we quote \cite{stat4},
``the model is differentiable in $\theta$ and that $\Theta_0$ and
$\Theta$ are (locally) equal to linear spaces''. Essentially, such
conditions imply that $\theta_{\min}$ follows an approximately Gaussian
distribution centered at the true $\theta$ value, which eventually
implies an approximate chi-square distribution for
$\Delta\chi^{2}_{min} (\theta_0)$.

In case I of our simulation, the best estimation of $\theta$ can be calculated directly from the number of events in each bin: \black{ $\theta_{\min}=\big[\sqrt{\sum^{n}_{i=1} N_i^2 /n}-1000\big]/15$ \,}\footnote{\black{Note that the above $\theta_{\min}$ can be closely approximated by $\big[(\sum^{n}_{i=1} N_i)/n-1000\big]/15$, which is indeed the exact maximum likelihood estimator for $\theta$ had we assumed that each count $N_i$ follows a Poisson distribution with mean $\mu_i(\theta)=1000+15\cdot \theta$.}}.The aforementioned regularity conditions are satisfied in this case, and the distribution of $\theta_{\min}$ and that of $\Delta\chi^{2}_{min}
(\theta_0)$ follow approximately the Gaussian and the chi-square distribution respectively 
as what Wilks theorem predicts. In the top two
panels of Fig.~\ref{fig:example}, we reconfirm this fact by comparing their
histograms based on the 100,000 MC samples (black shaded area) to the probability density 
function of the Gaussian and the chi-square distribution (blue long dash-dotted line). 
On the other hand, the full
parameter space in case II consists of two isolated points and clearly
violates the conditions required by Wilks theorem. Indeed, in case
II, the best estimation of $\theta$ is given by \begin{equation} \theta_{min} = \left\{ \begin{array}{ll}
      1 & \mbox{if $\chi^2(\theta=1) < \chi^2(\theta=-1)$} \nonumber\\
      -1 &\mbox{otherwise}
\end{array}\right. 
\end{equation} follows a Bernoulli distribution, 
and $\Delta\chi^{2}_{min} (\theta_0)$ follows a distribution 
quite different from a canonical chi-square distribution. Approximations to the actual
distributions of $\Delta\chi^{2}_{min} (\theta_0)$ and $\theta_{min}$ can be obtained from the 100,000 MC samples, 
and are shown (black shaded area) in the bottom two panels of Fig.~\ref{fig:example}. 
Further, an analytical approximation (red dash-dot-dotted line) to the distribution is derived in Appendix~A. 
The analytical calculation implies that, independent of whether the truth $\theta_0$ is $1$ or $-1$, 
the p-value~\footnote{The p-value at $t$ is defined to be the percentage of potential measurements 
that result in the same or a more extreme value of the test statistic, say $\Delta \chi^2_{min}$, 
than $t$.} corresponding to an observed value of $\Delta \chi^2_{min}(\theta_0)$, say $t$, is approximately given by 
\begin{equation}\label{eq:pvalue}
\text{p-value}(t)=P(\Delta \chi^2_{min}\left(\theta_0)\geq t\right) \approx  \frac{1}{2}-\frac{1}{2}\text{erf}\left(\frac{t+\overline{\Delta \chi^2}}{\sqrt{8\overline{\Delta \chi^2}}}\right)
\end{equation}
for any $t>0$; and the p-value is 1 for any $t\leq 0$.
Here erf is the Gaussian error function: $\text{erf}(x)=\frac{2}{\sqrt{\pi}}\int_0^x e^{-t^2}dt$. (We use the general notation $P(A)$ to denote the probability of an event $A$.)

The discussions above suggest that, when constructing confidence
intervals in special cases where conditions of Wilks theorem do not
hold (or that the user can not be sure if the conditions hold), the
regular threshold values (such as $t_\alpha=$ 1, 4, 9 mentioned
earlier) should not be taken for granted. Instead, alternative
thresholds based on MC or case-specific analytical approximations are
needed. We recommend using the MC method with a large MC sample size
whenever possible, because unlike other methods, it is guaranteed to
produce a valid confidence interval for $\theta$. We hereby review how to produce a valid 1-$\sigma$
(68.27\%) confidence interval for $\theta$ using the MC method~\cite{FC_stat}. This
method can easily be generalized to build confidence intervals of any level.
\begin{itemize} \item Having observed data $x=\{N_1,\cdots, N_n\}$,
  apply the following procedure to every $\theta$ in the parameter
  space $\Theta$ (fix one $\theta$ at a
  time): 
\begin{enumerate}
\item Calculate $\Delta \chi^2_{min}(\theta)^{x}$ with
  Eq.~\ref{eq:chisqmin} based on the observed data.
\item Simulate a large number of  MC samples, say $\{x^{(j)}\}_{j=1}^{T}$, where $x^{(j)}=\{N_1^{(j)},\cdots,N_n^{(j)}\}$ is generated from the model with true parameter value $\theta$.  For $j=1,\ldots,T$, calculate $\Delta \chi^2_{min}(\theta)^{(j)}$, that is  Eq.~\ref{eq:chisqmin} based on the $j$th MC sample 
$x^{(j)}$. 
This produces an empirical  distribution of the statistic
  $\Delta \chi^2_{min}(\theta)$.
\item Calculate the percentage of MC samples such that $\Delta \chi^2_{min}(\theta)^{(j)}
<\Delta \chi^2_{min}(\theta)^{x}$. Then $\theta$ is included in the 1-$\sigma$ 
confidence interval if and only if the percentage is smaller than 68.27\%. 
\end{enumerate}
\end{itemize}
One can easily check that p-values analytically obtained from 
Eq.~\ref{eq:pvalue}  for case II (basically the MC method) are 
consistent with the simulation results listed in Table~\ref{table:par}.

On a separate issue that was also emphasized in Ref.~\cite{FC_stat},
classical confidence intervals should not be confused with Bayesian
credible intervals. On one hand, the {\bf confidence-level} of a
confidence interval, say $\al$, is an evaluation of this interval
estimation procedure based on many {\bf potential} repetitions of the
experiment. More specifically, had the experiment been independently
repeated $100$ times, applying the estimation procedure to each would
result in $100$ intervals, and $\al$ represents the proportion of
these intervals that we expect to contain the true value of the
unknown parameter $\theta$. The level-$\al$ confidence interval
reported in practice is the result of applying such a procedure to the
data observed in the current experiment. On the other hand, a Bayesian
credible interval, say of {\bf credible-level} $\bl$, is a region in
the parameter space such that, given the {\bf observed} data, it
contains the true value of the unknown parameter with probability
$\bl$.
In general, an $\al$-level confidence interval does not coincide with
an $\al$-level Bayesian credible interval. In other words, if
$C_{\al}$ is an $\al$-level confidence interval built from the
observed data $x$, then it is generally inappropriate to give the
interpretation that $P(\theta\in C_{\al}|x)$ (the probability of 
true $\theta$ inside $C_{\al}$ given data $x$) is $\alpha$.
Nevertheless, in Appendix~\ref{sec:ap2}, we discuss when
 confidence intervals approximately match Bayesian credible
intervals. In the next section, we present a Bayesian approach to
the problem of determining neutrino mass hierarchy.

\section{A Bayesian Approach to Determine Neutrino Mass Hierarchy}\label{sec:bayes}

\begin{table*}
\begin{center}
\begin{tabular}{|c|c|c|c|c|c|c|c|c|}
\hline
$\alpha$ & 0.475 & 1 &1.281&1.645& 2 & 3 & 4 & 5   \\\hline
one-sided p-value: $p_{\alpha}$ & 31.74\% & 15.87\%& 10\%&5\% & 2.28\%& 0.13\%& 3.2e-5& 3.0e-7\\\hline
$\Delta \chi^2_{\alpha\sigma}$ & 1.53 & 3.33 & 4.39& 5.89 & 7.52& 13.29& 20.70& 30.04\\
\hline
\end{tabular}
\end{center}
\caption{\label{table:chi2} Tabulated results of $\Delta \chi^2_{\alpha\sigma}$. For a given $\alpha$, the one-sided p-value is $p_\alpha=P(Z\geq\alpha)$ (probability of Z $\geq$ than $\alpha$) where $Z$ stands for a standard Gaussian random variable. The corresponding $\Delta \chi^2$ value is given by $\Delta \chi^2_{\alpha\sigma}=-2\log(p_\alpha/(1-p_\alpha))$. }
\end{table*}

\subsection{Bayesian inference based on observed data} 
The MH determination problem is concerned with comparing two competing
models, NH and IH, having observed data $x$. The Bayesian approach to
the problem is based on the probabilities that each model is true
given $x$, namely, $P(NH|x)$ and $P(IH|x)=1-P(NH|x)$. 
(In general, we adopt the notation $P(A|B_1, \cdots, B_n)$ to
represent the probability of event $A$ given events $B_1, \cdots,
B_n$. Also, we use capital letters such as $S_1,\cdots, S_n$ and $T$
to denote random variables, and use small letters such as $s_1,\cdots,
s_n$ and $t$ to denote numbers inside the range of possible values of
the random variables. Then $P_{T|S_1,\cdots,S_n}(t|s_1,\cdots, s_n)$
denotes for the conditional probability density function (pdf)
or the conditional probability mass function
(pmf) 
given events $S_1,\cdots,S_n=s_1,\cdots,s_n$.  The subscript to $P$ is
often omitted when it is clear what random variable is being
considered.) Model NH will be preferred over IH if the odds
$r(x)=P(IH|x)/P(NH|x)<1$.  Moreover, the size of $r$ serves as an
easy-to-understand measure for the amount of certainty of this
preference. Alternatively, some people may feel more comfortable in
interpreting $P(NH|x)=1/(1+r(x))$ directly.

One can determine $P(NH|x)$ and $P(IH|x)$ within a
Bayesian framework as follows. Let the true value of MH
be either NH or IH, and let 
the counts $N_i$ follow a Gaussian distribution with
mean $\mu_i^{\text{MH}}(\theta,\eta_{\text{MH}})$ and standard
deviation $\sqrt{\mu_i^{\text{MH}}(\theta,\eta_{\text{MH}})}$ for
$i=1,\cdots,n$. Here, $\theta$ is the parameter of interest, and
$\eta_{\text{MH}}$ denotes other unknown nuisance parameter(s). Here,
a subscript accompanies $\eta$ to emphasize that the nuisance
parameter is allowed to have different interpretations and behavior
under the two hypotheses. (We will omit this subscript whenever there
is no possibility of confusion.)
If prior knowledge is available for $\theta$ and $\eta$, then
they should be elicited to form prior distributions,
$P(\theta,\eta|MH)$ for MH=IH, NH.  Sometimes, it is reasonable to
assume that the parameters $\theta$ and $\eta$ are independent
conditional on MH, hence $P(\theta, \eta|
MH)=P(\eta|MH)P(\theta|\eta,MH)=P(\eta|MH)P(\theta|MH)$.

Specific to the MH problem at hand, under NH (IH), previous
  knowledge (e.g., from \cite{PDG}) suggests that a sensible prior for
  $\theta$ would be a Gaussian with mean $2.43\times10^{-3}$ eV$^2$
  ($-2.43\times10^{-3}$ eV$^2$) and standard deviation
  $0.13\times10^{-3}$ eV$^2$. Since the hypotheses being tested are
  $\text{NH}: \theta \in \Theta_{NH}=(0,\infty)$ versus $\text{IH}:
  \theta \in \Theta_{IH}=(-\infty, 0)$, $P(\theta|NH)$ and $P(\theta|IH)$
  are specified to be the truncated version of the above Gaussian
  distributions supported within $\Theta_{NH}$ and $\Theta_{IH}$,
  respectively.
  Nevertheless, in our Bayesian model, $P(\theta\in\Theta_{IH}|NH)$ and
  $P(\theta\in\Theta_{NH}|IH)$ based on the Gaussian prior are so tiny that they
  will yield the same numerical results as the truncated version.
Similar choice can be made for $P(\eta|MH)$.

According to Bayes' theorem, we have \begin{eqnarray}\label{eq:b2}
  P(NH|x) &=& \frac{P(x|NH)\cdot P(NH)}{P(x)}  \\
  &=& \frac{P(x|NH)\cdot P(NH)}{P(x|NH)\cdot P(NH) + P(x|IH)\cdot
    P(IH)}\,. \nonumber
\end{eqnarray}
Here, $P(NH)$ and $P(IH)=1-P(NH)$ should reflect one's knowledge in NH
and IH prior to the experiment.  In the MH problem, it is reasonable
to assume that NH and IH are equally likely, that is $P(NH) = P(IH) =
50\%$. We will make this assumption throughout the
paper. Consequently, Eq.~\ref{eq:b2} reduces to
\begin{equation}\label{eq:b21}
  P(NH|x) = \frac{P(x|NH)}{P(x|NH) + P(x|IH)}.
\end{equation}
Based on probability theory, $P(x|MH)$, i.e. the likelihood of model
MH, is a ``weighted average" of $P(x|\ttheta,\eta,MH)$ over all
possible values of $(\ttheta, \eta)$:
\begin{equation}\label{eq:b3}
  \begin{split}
  &P(x|MH) \\
  =&  \int_{H_{\text{MH}}}\int_{\Theta_{\text{MH}}} P(\eta|MH)  P(\ttheta|\eta, MH) P(x|\ttheta,\eta,MH) d \ttheta d\eta\,,
\end{split} \end{equation} in which $H_{\text{MH}}$ represents the
phase space of nuisance parameter $\eta$ given the choice of MH.
Further, 
under the assumption that $\theta$ and $\eta$ are independent,
Eq.~\ref{eq:b3} is reduced to:
\begin{equation}\label{eq:b4}
  \begin{split}
    &P(x|MH) \\
    =& \int_{H_{\text{MH}}}\int_{\Theta_{\text{MH}}} P(\eta|MH) P(\ttheta|
    MH)P(x|\ttheta,\eta,MH)d \ttheta d\eta
    \,.
\end{split}  
\end{equation} 

In practice, the integral in Eq.~\ref{eq:b3} is often analytically
intractable, but can be approximated using MC methods. Using
a basic MC scheme, first, a large number of samples
$\{({\ttheta}^{(j)}, \eta^{(j)}), j=1,\ldots,T\}$ are randomly
generated from the prior distribution $P(\ttheta,\eta| MH)$.  Then for
the observed data $x$, obtain $\hat{P}_T(x|MH):=T^{-1}\sum_{j=1}^T
P(x|{\ttheta}^{(j)},\eta^{(j)},MH)$. As the MC size $T$
increases, the estimator $\hat{P}_T(x|MH)$ will have probability
approaching $1$ of being arbitrarily close to the true $P(x|MH)$.
Note that there exist much more efficient MC algorithms, such
as importance sampling algorithms, that require smaller, more
affordable $T$ for the resulting estimators to achieve the same amount
of accuracy as that of the basic MC scheme. Interesting
readers are pointed to \cite{bayes} for further details and
references.




There also exist (relatively crude) approximations to $P(x|MH)$ in
Eq.~\ref{eq:b3} that avoid the intense computation in the MC
approach.  A most commonly used  one is the one on which a
popular model selection criteria, the Bayesian information criterion
(BIC) is based.
This approximation is often presented in terms of an approximation to
a one-to-one transformation of $P(x|NH)$,
namely \begin{equation}\label{eq:chi2}\Delta \chi^2(x)\equiv -2\log r(x)=-2\log
  \left(P(IH|x)/P(NH|x)\right)\,.
\end{equation}
Denote
\[\T_{\text{MH}}(x) \equiv -2 \log
\{\max_{\ttheta,\eta} P(x|\ttheta,\eta,MH) P(\eta|MH) P(\ttheta| MH)
\}\,,\] where the maximum is taken over
$(\theta,\eta)\in\Theta_{\text{MH}}\times H_{\text{MH}}$ and
\begin{equation}\label{eq:dchi}\Delta\T(x) \equiv \T_{IH}(x)-\T_{NH}(x)\,.\end{equation}
Then if the sample size $\sum_i N_i$ is large, and $\eta_{NH}$
and $\eta_{IH}$ are of the same dimension,
then \begin{equation}\label{eq:bic}\begin{split}
    \Delta \chi^2(x)&= 2\log P(x|NH)-2\log P(x|IH)\\
    &\approx
\Delta \T(x)\,. 
\end{split}
\end{equation}
Here, the equality follows from Eq.~\ref{eq:b21} and
the approximation is supported by a crude Taylor expansion around the
maximum likelihood estimator for the parameters. There are other
approximations that follow the same line, that are more accurate but
also computationally more demanding. See \cite{bayes} for details.
  
One remark should be made regarding $\Delta \T$, as it is closely
related to a commonly used test statistic in the classical testing
procedure. Indeed, if the truncated Gaussian priors mentioned earlier
are assigned for $\theta$ and a Gaussian prior with mean $\eta_0$ and
standard deviation $\delta \eta$ is assigned for $\eta$ under both NH
and IH, then according to the definition of $\chi^2$ in
Eq.~\ref{eq:chi2def}, we have
\begin{equation}\label{eq:relate}
  \DC \equiv \chi^2(\hat{\theta}',\hat{\eta}')-\chi^2(\hat{\theta},\hat{\eta})=\Delta \T-\sum_{i=1}^n \log\frac{\mu_i(\hat{\theta}',\hat{\eta}')}{\mu_i(\hat{\theta},\hat{\eta})},
\end{equation}
where $(\hat{\theta},\hat{\eta})$ and $(\hat{\theta}',\hat{\eta}')$
denote maximizers of
\[
P(x|\ttheta,\eta,MH)P(\eta|MH) P(\ttheta| MH)\,,\;\;\;\;\text{MH=NH,
  IH.}
\] within their respective range. (Note that $\DC$ is essentially an
alternative version of $\Delta \chi^2_{\min}$ in Eq.~\ref{eq:chisqmin},
bearing some technical difference only.) Here, the term
$\sum_{i=1}^n
\log\frac{\mu_i(\hat{\theta}',\hat{\eta}')}{\mu_i(\hat{\theta},\hat{\eta})}$
is the result of the normalization factor
(e.g. $(2\pi\sigma^2)^{-\frac{1}{2}}$) of the Gaussian pdf, and is in
general small compared to $\Delta \T$.
In the classical testing procedure, the observed value of $\DC$ will
be compared to its parent distribution to get a p-value. Whereas the
Bayesian approach described in this section directly interprets the
value of $\Delta \T$, by transforming it to either the odds ratio
between NH and IH, $r(x) = e^{-\Delta \chi^2(x)/2} \approx e^{-\Delta
  \T(x)/2}$, or the probability of NH,
\begin{eqnarray}\label{eq:cal}
  P(NH|x)  &=& \frac{1}{1+r(x)} \nonumber \\
	&=& \frac{1}{1+e^{-\Delta \chi^2(x)/2}} \approx \frac{1}{1+e^{-\Delta \T(x)/2}}  \,,
\end{eqnarray}
and similarly, the probability of IH,
\begin{eqnarray}\label{eq:cal1}
  P(IH|x)  &=& \frac{r(x)}{1+r(x)} \nonumber \\
	&=& \frac{e^{-\Delta \chi^2(x)/2} }{1+e^{-\Delta \chi^2(x)/2}} \approx \frac{e^{-\Delta \T(x)/2}}{1+e^{-\Delta \T(x)/2}}  \,.
\end{eqnarray}


\subsection{Sensitivity of experiments}
So far, we described the Bayesian procedure for testing the two
hypotheses, NH and IH, given observed data
$x=\{N_1,\cdots,N_n\}$. Reasoning backwards, foreseeing what analysis
will be done after data collection allows us to address the question
that, before data is collected from a proposed experiment, how confidently
 do we expect it to be able to distinguish the two hypotheses NH
and IH.  We loosely refer to such an ability as ``sensitivity" of the
experiment.
There could be many ways to define sensitivity, and we list a few
below. In practice, evaluating a proposed experiment using one or
several of these sensitivity criteria provides views from different
angles of the potential return from the experiment.

Note that sensitivity depends on the underlying true model as well as
future experimental results generated from this model. For example, if
NH is true, then we have a population of potential experimental
results $x\sim P(x|NH)=\int\int
P(x|\theta,\eta,NH)P(\theta,\eta|NH)d\theta d\eta$. And each potential
$x$ is associated with a posterior probability $P(NH|x)$. Then one
could evaluate the ability of an experiment to confirm NH when it is
truly the underlying model by looking at the distribution of
$P(NH|x)$. The most typical numerical summaries of this distribution
include its mean, quantiles and tail probabilities, all of which can
be used to address sensitivity. 

\black{Below we officially develop metrics for sensitivity under the assumption that
NH is true. Note that these metrics can be similarly defined when IH is
true. 
}
\begin{enumerate}

\item The average posterior probability of NH is given by
  \begin{eqnarray}\label{eq:avep}
    \overline{P}_{T=NH}^{NH} &=& \int P(NH|x) \, P(x|NH)  \,dx\\
    &=& \int_{-\infty}^{\infty} \frac{1}{1+e^{-\Delta \chi^2/2}}\,  P(x|NH)  \,dx\nonumber\\
    &=& \int_{-\infty}^{\infty} \frac{1}{1+e^{-\Delta \chi^2/2}} \,  P(\Delta \chi^2|NH)  \,d\Delta \chi^2\,.\nonumber
  \end{eqnarray}
  Note that the first integral above involves calculating an $N$-dim
  integral, and the last one is of $1$-dim only. The latter is much
  easier to obtain, an example of which will be presented in the next
  section. 


\item The fraction of measurements $x$ that favor NH, i.e., the 
fraction of $x$ such that
  $P(NH|x)>.5$, is given by
\begin{eqnarray}\label{eq:PC0}
  F_{T=NH} &=& \int_{\{x: P(NH|x)>.5\}}P(x|NH) dx \nonumber\\
  &=& \int_0^{\infty} P(\Delta\chi^2|NH) d\Delta \chi^2\,.
\end{eqnarray}
Here, ``F'' and the subscript ``$T=NH$" stand for fraction and the NH assumption, respectively.  

If NH is the correct hypothesis, then a good experiment should have a
high probability of producing data that not only favors NH but indeed
provides substantial evidence for NH.  Hence, it is useful to
generalize the term in Eq.~\ref{eq:PC0} to gauge the chance of
$P(NH|x)>1-p$ for any threshold value $1-p$ of interest.
In particular, physicists are familiar with thresholds associated with the
so-called $\alpha \sigma$ level, with one-sided $\alpha\sigma$ corresponding to
$1-p_\alpha=1-P(Z\geq \alpha)$ for a standard Gaussian random variable
$Z\;$~\footnote{Another commonly used term is two-sided $\alpha\sigma$ 
which corresponds to $1-P(|Z|\geq\alpha)$. }. 
Accordingly, define
\begin{eqnarray}\label{eq:PC}
  F_{T=NH}^{\alpha\sigma} &=& \int_{\{x: P(NH|x)>1-p_{\alpha}\}}P(x|NH) dx \nonumber\\
  &=& \int_{\Delta \chi^2_{\alpha\sigma}}^{\infty} P(\Delta\chi^2|NH) d\Delta \chi^2.
\end{eqnarray}
A list of common $\alpha\sigma$ values, the corresponding $p_\alpha$,
as well as $\Delta \chi^2_{\alpha\sigma}=-2\log(p_\alpha/(1-p_\alpha))$ are listed
in Table.~\ref{table:chi2}.

\item In addition, probability intervals (PI) for
  $P(NH|x)$ also provide useful information. \black{ For example, a 90\% PI is denoted by $(P^{90\%}_{T=NH}, 1)$, where
  $P^{90\%}_{T=NH}$ is the 100-90=10th percentile of $P(NH|x)$. That is, had NH been the truth, 90\% of the
  potential data would yield $P(NH|x)$ larger than
  $P^{90\%}_{T=NH}$. }
\end{enumerate}
All the above criteria reflect the capability of the
experiment \black{to} distinguish the two competing hypotheses, and they
convey different messages.

\black{Finally, to get a complete picture of the
sensitivity of an experiment, one should also obtain the above
metrics under the assumption that IH is the underlying true model. The sensitivity scores under metrics 2 and 3 can be shown to depend on the underlying true model. For example, we experimented with simple examples (not shown) and observed that in general $F_{T=NH} \neq F_{T=IH}$. 
Whereas for metric 1, we have $\overline{P}_{T=NH}^{NH}=\overline{P}_{T=IH}^{IH}$ as long as equal prior probabilities~\footnote{We acknowledge the referee for pointing out this important relation.}, $P(NH)=P(IH)$, were assigned to the two models. This is because \[\begin{split}
\overline{P}_{T=NH}^{NH}&-\overline{P}_{T=IH}^{IH}\\&=\int  P(NH|x) P(x|NH)-  P(IH|x) P(x|IH)d x 
\\
&= \int \frac{P^2(x|NH) P(NH)-P^2(x|IH) P(IH)}{P(x|NH) P(NH)+P(x|IH) P(IH)}  d x\\
&=\int P(x|NH)-P(x|IH)  d x=0\,.
\end{split}\]
}

In the next section, we use an example to show how one can easily
calculate the posterior probability and the sensitivity measurements
introduced above. We also contrast the resulting sensitivity
measurements to a commonly used quantity that is known as
``$\overline{\Delta \chi^2}$ of Asimov data set''.

\section{Illustration of the Bayesian Approach in a constrained parameter space}\label{sec:example}


\begin{figure}[]
\centering
\includegraphics[width=90mm]{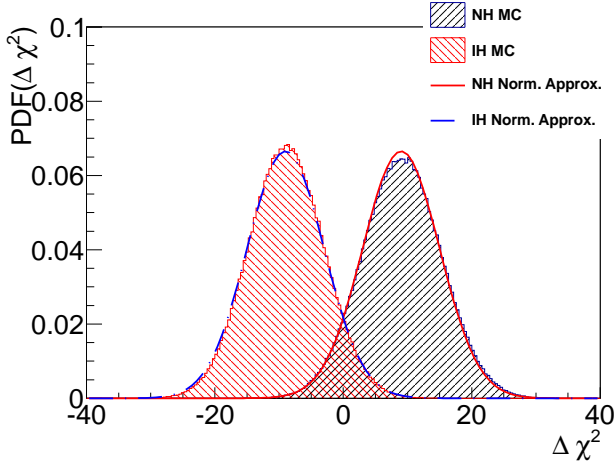}
\caption{(color online) The probability density functions $P(\Delta\chi^2|NH)$ and 
$P(\Delta\chi^2|IH)$ in the Bernoulli model are
shown as the solid and dotted lines, respectively. 
The $|\overline{\Delta\chi^{2}}|$ is assumed to be 9.}
\label{fig:chi2}
\end{figure}

In this section, we consider a situation where $\theta$ can only take on
two possible values, $1$ and $-1$, which correspond to the hypotheses
NH and IH respectively. This simplified setting is motivated by
the fact that existing measurements of $|\theta|=M^2_{32}$ are very
accurate at around $2.43\times10^{-3}$ eV$^2$, and we simply
denote this value to $1$ for clarity of presentation.  It is
a special case of the Bayesian treatment in the previous section, where
$P(\theta|NH)$ and $P(\theta|IH)$ are assigned degenerate
distributions at $1$ and $-1$ respectively. That is,
$P(\theta=1|NH)=P(\theta=-1|IH)=1$. Further, there is no nuisance
parameter $\eta$. As a result, the expected bin counts will be denoted
by $\mu_{i}^{NH}=\mu(1)$ and $\mu_{i}^{IH}=\mu(-1)$ respectively.

Below, we showcase numerical calculations of various sensitivity
criteria for this example.
In particular,  we introduce approximations that are simple functions of a term commonly known as  ``$\overline{\Delta \chi^2}$ of 
Asimov data set''
in the physics 
literature. According to the definition in Ref.~\cite{Cowan}, 
``the Asimov data set'' under hypothesis MH is given by
  $x^{MH}=(\mu_1^{MH},\cdots,\mu_N^{MH})$, where
  $\mu_i^{MH}=\mu_i(\theta_0^{MH},\eta_0^{MH})$ and $(\theta_0^{MH},
  \eta_0^{MH})=\arg\max_{(\theta,\eta)}P(\theta,\eta|MH)$ is the prior
  mode under MH. In words, the Asimov data set is the most typical
  data set under the most likely parameter values based on prior
  knowledge subject to the given model.  

Interestingly, $\overline{\Delta \chi^2}$ is itself often used as a measure of sensitivity. Here, we'll contrast the typical usage of $\overline{\Delta \chi^2}$ to that of the sensitivity criteria developed in the previous section. More accurate evaluations of these sensitivity criteria are also attainable via 
MC methods.


Suppose that the proposed experiment will collect enough data such
that the expected counts under NH and IH are much larger than the
difference between them: $ \mu_{i}^{NH} \sim \mu_{i}^{iH}
>>|\mu_{i}^{NH} -\mu_{i}^{iH}|$.
Using the notations introduced in Sec.~\ref{sec:parameter}, if the
nature is NH, then the observed counts $N_i$ can be represented as
\begin{equation}
  N_{i}=\mu_{i}^{NH}+  \sqrt{\mu_{i}^{NH}}\cdot g_{i},\label{eq:true-Gaussian}
\end{equation}
where $g_1, \cdots, g_n$ are mutually independent standard Gaussian
random variables. Then, 
the statistic $\Delta \chi^2$ of 
Eq.~\ref{eq:chi2} becomes 
\begin{eqnarray}
\Delta\chi^{2}_{T=NH} 
&=& \sum_{i=1}^{n} \frac{\left(\mu_{i}^{NH}-\mu_{i}^{IH}\right)^{2}}{\mu_i^{IH}} \nonumber \\ 
&+& \sum_{i=1}^{n}  2\frac{\left(\mu_{i}^{NH}-\mu_{i}^{IH}\right)\sqrt{\mu_i^{NH}}g_i}{\mu_i^{IH}} \nonumber \\ 
&+&  \sum_{i=1}^{n} \frac{\mu_i^{NH}-\mu_i^{IH}}{\mu_i^{IH}}g_i^2\nonumber \\
&-&  \sum_{i=1}^{n} \log (1+\frac{\mu_i^{NH}-\mu_i^{IH}}{\mu_i^{IH}})\,.
\label{eq:delta-chi-Gaussian}
\end{eqnarray}
Here, the subscript $T=NH$ indicates that the nature is NH.
Since $\mu_{i}^{iH} >>|\mu_{i}^{NH} -\mu_{i}^{iH}|$, the summation of
the last two terms in Eq.~\ref{eq:delta-chi-Gaussian} is negligible as
it is approximately $ \sum_{i=1}^{n}
\frac{\mu_i^{NH}-\mu_i^{IH}}{\mu_i^{IH}} \cdot (g_i^2-1)$ by a Taylor
expansion of the last term. Therefore, $\Delta\chi^{2}_{T=NH} $
follows a Gaussian distribution,
with mean and standard deviation:
\begin{equation}
\begin{cases}
\begin{array}{lll}
\overline{\Delta\chi^{2}} &\equiv & \sum_{i=1}^{n}\frac{\left(\mu_{i}^{NH}-\mu_{i}^{IH}\right)^{2}}{\mu_i^{IH}}\\
\sigma_{\Delta\chi^{2}} & \equiv & 2\sqrt{\sum_{i=1}^{n}\frac{\left(\mu_{i}^{NH}-\mu_{i}^{IH}\right)^{2}\cdot \mu_i^{NH}}{(\mu_i^{IH})^2}} \\
&=&2\sqrt{\sum_{i=1}^{n}\left(\frac{\left(\mu_{i}^{NH}-\mu_{i}^{IH}\right)^{2}}{\mu_i^{IH}} + \frac{\left(\mu_{i}^{NH}-\mu_{i}^{IH}\right)^{3}}{(\mu_i^{IH})^2} \right)} \\
& \approx& 2\sqrt{\overline{\Delta\chi^{2}}}
\end{array}\end{cases}. \label{eq:chi-bar-rms1}
\end{equation}
In the last step, since $\mu_i^{NH} - \mu_i^{IH} << \mu_i^{NH} \sim
\mu_i^{IH}$, we further neglect the term
$\frac{\left(\mu_{i}^{NH}-\mu_{i}^{IH}\right)^{3}}{(\mu_i^{IH})^2}$.
Similarly, it is straightforward to show that when nature is IH,
$\Delta\chi^{2}_{T=IH}$ would follow an approximate Gaussian
distribution with mean $=-\overline{\Delta\chi^{2}}$ and standard
deviation $\sigma_{\Delta\chi^{2}}$. \black{In fact, when IH is true, $\overline{\Delta\chi^{2}_{IH}} = -\sum_{i=1}^{n}\frac{\left(\mu_{i}^{NH}-\mu_{i}^{IH}\right)^{2}}{\mu_i^{NH}} \approx -\overline{\Delta\chi^{2}}$.}


To see how the above approximation works, we look at the
  example in Sec.~\ref{sec:parameter}, where $\overline{\Delta \chi^2}
  \approx 9$.  Fig.~\ref{fig:chi2} shows histograms (shaded area)
  based on large MC samples of $\Delta\chi^{2}$ under NH and
  IH respectively.
They agree very well with the analytical approximation (dashed lines)
in Eq.~\ref{eq:chi-bar-rms1}.

Now, we are ready to calculate (1) the probability of a hypothesis
post data collection, and (2) various measurements of sensitivity for
an experiment concerning potential data generated from
it.

First, given observed data $x=(N_1,\cdots,N_n)$, the probability
$P(NH|x)$ can be directly calculated from Eq.~\ref{eq:b2}.  Let $G(t;
m,\sigma)=\frac{1}{\sqrt{2\pi} \cdot \sigma}
e^{-\frac{(t-m)^2}{2\sigma^2}}$ denote the pdf of a Gaussian random
variable with mean $m$ and standard deviation $\sigma$, evaluated at
$t$, then 
\[\begin{split} P(NH|x)&= \frac{P(x|NH)\cdot P(NH)}{P(x|NH)\cdot P(NH) + P(x|IH)\cdot    P(IH)}\\
  &=\frac{\Pi_i G(N_i;\mu^{NH}_i,\sqrt{\mu^{NH}_i})}{\Pi_i G(N_i;\mu^{NH}_i,\sqrt{\mu^{NH}_i})+\Pi_i G(N_i;\mu^{IH}_i,\sqrt{\mu^{IH}_i})}\\
  &= \frac{1}{1+e^{-\Delta \chi^2(x)/2}}
\end{split}
\]
where 
\[\Delta \chi^2(x)=\sum_{i=1}^n\left[
  \log\frac{\mu_i^{IH}}{\mu_i^{NH}}+
  \frac{\left(N_i-\mu_i^{IH}\right)^2}{\mu_i^{IH}}-\frac{\left(N_i-\mu_i^{NH}\right)^2}{\mu_i^{NH}}
\right]\,.\]
We mention that, if one reduces the full
data $x$ to its function $\Delta \chi^2(x)$, then calculating
$P(NH|\Delta \chi^2)$ based on our approximation in
Eq.~\ref{eq:chi-bar-rms1} will recover $P(NH|x)$:
\begin{widetext}
   \begin{eqnarray}\label{eq:e1}
  P(NH|\Delta \chi^2)  &=& \frac{P(\Delta \chi^2|NH) \cdot P(NH)}{P(\Delta \chi^2)} = \frac{P(\Delta \chi^2|NH) }{P(\Delta \chi^2|NH)  + P(\Delta \chi^2|IH) } \nonumber \\ 
     &=& \frac{G\left( \Delta \chi^2;\overline{\Delta \chi^2},2\sqrt{\overline{\Delta \chi^2}} \right)}
     {G\left(\Delta \chi^2;\overline{\Delta \chi^2},2\sqrt{\overline{\Delta \chi^2}} \right) + G\left(\Delta \chi^2;-\overline{\Delta \chi^2},2\sqrt{\overline{\Delta \chi^2}} \right)} 
     = \frac{1}{1+e^{-\Delta \chi^2/2}}.
 \end{eqnarray}
 \end{widetext}

\begin{figure*}
  \centering
\includegraphics[width=150mm]{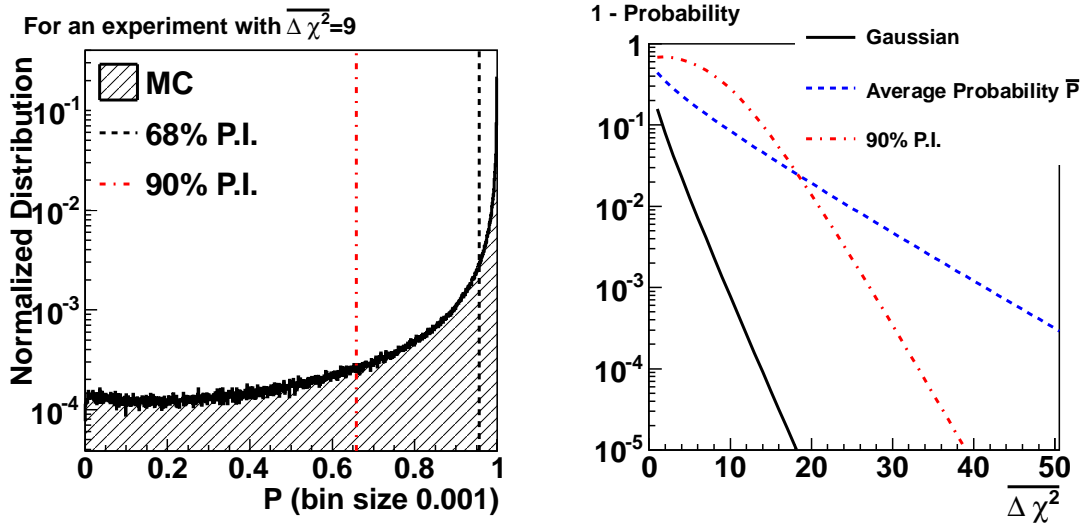}
\caption{(color online) The left panel shows the distribution of
  $P(NH|x)=P(NH|\Delta\chi^2)$ over the population of potential data
  $x$ that arises from an experiment with $\overline{\Delta \chi^2}=9$
  where the truth is NH. The mean of this distribution is 90.14\%.
  Lower bound of the 68\% and 90\% probability intervals are
  plotted. That is, 68\% (90\%) of the data $x$ would yield a
  $P(NH|x)$ that falls to the right of the dash-dotted (dashed)
  line. These two lines are also commonly referred as the \black{32th and the
  10th} percentile.  The right panel plots several sensitivity metrics
  (subtracted from $1$ for clarity), against $\overline{\Delta
    \chi^2}$ that ranges from 1 to 50. Note that all the lines are
  decreasing because higher values of $\overline{\Delta \chi^2}$
  corresponds to more sensitive experiments.  This is done for three
  different criteria: the Gaussian interpretation (derived from 
  the one-sided p-value with one degree of freedom),
  $\overline{P}$ and $P^{90\%}_{T=NH}$. The Gaussian interpretation is
  seen to be over-optimistic in describing the ability of the
  experiment to differentiate the two hypotheses. }
\label{fig:dis}
\end{figure*}


Next, 
we evaluate various sensitivity metrics of a future experiment,
using again the Gaussian distribution for $\Delta\chi^2$ in
Eq.~\ref{eq:chi-bar-rms1}:
\begin{widetext}
\begin{eqnarray}\label{eq:e2}
  \overline{P}_{T=NH}^{NH}
  &\approx& \int_{-\infty}^{\infty} 
  \frac{1}{1+e^{-t/2}} \,G\left(t,\overline{\Delta \chi^2},2\sqrt{\overline{\Delta \chi^2}} \right) d t
  \equiv \overline{P}(\overline{\Delta \chi^2})\,,\\
  F_{T=NH} &\approx&  \int_0^{\infty} G\left(t;\overline{\Delta \chi^2},
    2\sqrt{\overline{\Delta \chi^2}}\right) dt
  = \frac{1}{2} \left(1 + {\rm \text{erf}}\left(\sqrt{\frac{\overline{\Delta \chi^2}}{8}}\right)\right), \\
  F^{\alpha\sigma}_{T=NH} &\approx& \int_{\Delta \chi^2_{\alpha\sigma}}^{\infty} G\left(t;\overline{\Delta \chi^2},
    2\sqrt{\overline{\Delta \chi^2}}\right) dt
  =\frac{1}{2}\left(1+{\rm \text{erf}}\left(\frac{\overline{\Delta \chi^2}-\Delta \chi^2_{\alpha\sigma}}{\sqrt{8\overline{\Delta \chi^2}}}\right)\right)\,,\\
  P^{\A \%}_{T=NH}&\approx& 1\bigg/\left(1+e^{-\frac{1}{2}\left(\overline{\Delta \chi^2}-2 z^*_{\A} \sqrt{\overline{\Delta \chi^2}}\right)}\right)\,.\label{eq:elast}
\end{eqnarray}
\end{widetext}
\black{In Eq.~\ref{eq:e2} above, $\overline{P}_{T=NH}^{NH}$ was approximated by
$\overline{P}(\overline{\Delta \chi^2})$, which is a function of
$\overline{\Delta \chi^2}$
only. 
In Eq.~\ref{eq:elast}, $z^*_{\A}$ represents the $\A$th percentile of a
standard Gaussian distribution, hence $\overline{\Delta \chi^2}-2 z^*_{\A} \sqrt{\overline{\Delta \chi^2}}$ is the (100-A)th percentile of $\Delta \chi^2$ according to the Gaussian approximation in  Eq.~\ref{eq:chi-bar-rms1}. Since $P(NH|\Delta \chi^2)=1\big/(1+e^{-\Delta \chi^2/2})$ is increasing in $\Delta \chi^2$, this means that the righthand side of Eq.~\ref{eq:elast} is the (100-A)th percentile of $P(NH|\Delta \chi^2)$, which serves as the lower bound of the A\% PI proposed in the previous section.}
\black{ In Table.~\ref{table:pi}, we list 
$z^*_{\A}$ for a few typical choices of probability
  intervals, assuming that the nature is NH. }
\begin{table*}
\begin{center}
\begin{tabular}{|c|c|c|c|c|}
\hline
$A\%$ & 68\% & 90\% & 95\%  & 99\%  \\\hline
\black{Gaussian Percentile $z^*_A$} & 0.468 & 1.282 & 1.645 & 2.326  \\\hline
\end{tabular}
\end{center}
\caption{\label{table:pi} Tabulated $\Delta \chi^2_{PI}$ values for a few typical choice of probability 
intervals, assuming that nature is NH. }
\end{table*}

For the example experiment used in the simulation of
section~\ref{sec:parameter}, its $\overline{\Delta\chi^{2}} = 9$.  Had
one followed common practice that directly compares
$\sqrt{\overline{\Delta\chi^{2}}}$ to the quantiles of a Gaussian
distribution, one would report the ``specificity'' of the experiment
to be 99.87\% (1 - ``one-sided p-value''). 
In contrast, we obtained various sensitivity metrics for
the experiment according to Eq.~\ref{eq:e2}-\ref{eq:elast}, and listed
them in Table~\ref{table:summary}.  First, assuming the ``Asimov data set''
is observed, we have
$P(NH|x^{NH}) \approx P(IH|x^{IH}) \approx P(NH|\Delta \chi^2=9) =
98.90\%$.  \black{Secondly, we calculated $\overline{P}_{T=NH}^{NH}=
\overline{P}_{T=IH}^{IH} \approx \overline{P}(\overline{\Delta
  \chi^2}=9)= 90.14\%$.}  That is, the average posterior
  probability for NH (or IH) when it is indeed the correct hypothesis
  is only about $90\%$, which is much lower than its Asimov
  counterpart of $P(NH|\Delta \chi^2=9)=98.90\%$. Thirdly, the fraction $F_{T=NH}
= 93.32\%$ of potential data sets would yield a $\Delta \chi^2$ that favors
NH. And to contrast with the Gaussian interpretation, we
calculated that only $F^{3\sigma}_{T=NH}= 23.73\%$ of potential data sets
would yield a $\Delta \chi^2$ above 9, or say, yield an evidence as
strong as $P(NH|x)\geq 99.87\%$.  Further, the left panel of
Fig.~\ref{fig:dis} displays the distribution (vertical axis in log
scale) of $P(NH|x)=P(NH|\Delta\chi^2)$. The two vertical dashed lines
show that $68\%$ of potential data sets will result in $P(NH|x)>95.67\%$,
whereas $90\%$ of potential data sets will result in $P(NH|x)>65.79\%$.

\begin{table*}
\begin{center}
\begin{tabular}{|c|c|c|c|c|c|c|c|}
  \hline
  Symbol & $\overline{P}$ & & $P(NH|x)$ & $F_{T=NH}$ 
  &$F^{3\sigma}_{T=NH}$ & $P^{68\%}_{T=NH}$ &  $P^{90\%}_{T=NH}$ \\
 Description &Average          &  Gaussian Interpretation& Asimov data set & $\Delta \chi^2>0$ & $P>99.87\%$   & 68\% P.I. & 90\% P.I. \\\hline
  Sensitivity Metric &90.14\%&99.87\% & 98.90\% & 93.32\% &23.73\%  &95.67\% & 65.79\% \\
  \hline
\end{tabular}
\end{center}
\caption{\label{table:summary} Sensitivity metrics for an experiment with $\overline{\Delta \chi^2}=9$. }
\end{table*}

Moving forward from a fixed $\overline{\Delta \chi^2}$ value, we next
study how the various sensitivity metrics compare to each other for
experiments with different $\overline{\Delta \chi^2}$ values. The
right panel of Fig.~\ref{fig:dis} displays
the lower bound of the 90\% probability interval $P^{90\%}_{T=NH}$,
the average probability $\black{\overline{P}_{T=NH}^{NH}}$, and the
Gaussian interpretation based on one-sided p-value as functions of
$\overline{\Delta \chi^2}$. Note that we plotted $1$ minus the
aforementioned metrics in order to zoom in the high probability
regions. 
Interestingly, the line of average probability $\overline{P}$ yields a
higher value than the lower bound of 90\% P.I. for $\overline{\Delta
  \chi^2}<\sim18$, and yields a lower value than the lower bound of
90\% P.I. for $\overline{\Delta \chi^2}>\sim18$.  Such behavior is
natural given the definition of each curve.  Nevertheless, both curves
are much higher than the Gaussian interpretation, suggesting
that the Gaussian interpretation is over-optimistic in
describing the ability of an experiment to differentiate NH and IH.

\section{Discussions}\label{sec:discussion}
A couple of comments should be made regarding the $\sqrt{\overline{\Delta
    \chi^2}}$ representation for sensitivity in determining the
MH. 
\begin{enumerate}
\item We have seen that the distribution of the best estimator of
$\theta=\Delta m^2_{32}$ is closer to a Bernoulli distribution than to a
Gaussian distribution. Therefore, Wilks' theorem is not
applicable, and direct interpretation of $\sqrt{\Delta\chi_{min}^{2}}$
as the number of $\sigma$ in the Gaussian approximation leads to
incorrect confidence intervals. We provided an analytical formula
(Eq.~\ref{eq:pvalue}) for confidence interval in an ideal Bernoulli
case, which can be used to generate approximate confidence intervals
for similar cases. For more general cases, a full MC simulation is
needed to construct confidence intervals, as advocated in
Ref.~\cite{FC_stat}. 
\item Even if a confidence interval for $\Delta m^2_{32}$ is  constructed correctly, its confidence level can not be directly 
interpreted as how
much the current measurement would favor the NH (IH) against the
other. Despite possible agreement between confidence intervals and
Bayesian credible intervals under certain circumstances as discussed
in Appendix.~\ref{sec:ap2}, such agreement does not apply to the current MH problem where there are strong constraints imposed on $M^2_{32}$. 
\end{enumerate}


Additional comments should be made regarding the
  Bayesian approach. \begin{enumerate}
  
  
  \item In principle, results from different experiments
  can be combined within the Bayesian framework. One example 
  can be found in Ref.~\cite{Berg}, in which a Bayesian 
  method was applied to constrain $\theta_{13}$
  and CP phase $\delta$ with existing experimental data.
  Regarding to the MH, results from different experiments
  can be combined through the integral in Eq.~\ref{eq:b3}.  
  Specifically, one can integrate over the
  nuisance parameters regarding experimental systematic
  uncertainties, while leaving nuisance parameters regarding the
  relevant neutrino masses and mixing parameters unintegrated. 
 For example, suppose there are two independently conducted experiments, labeled by $j=1, 2$, and that their respective observed data $x_j$ corresponds to the model $P(x_j|\theta,\eta^*,\eta_j,MH)$ under MH=NH or IH. Here the vector of nuisance parameter $\eta$ in experiment $j$ is separated into two pieces $\eta^*$ and $\eta_j$, where $\eta_j$ is unique to the experiment and $\eta^*$ is common to both experiments. Of course, $\theta$ is the parameter of interest and hence always common to both. Then, it would be useful for the different experiments to not only present  $\Delta \chi^2$  (Eq.~\ref{eq:chi2}), but to also present
  \[P(x_j|\theta,\eta^*,MH)=\int P(x_j|\theta,\eta^*,\eta_j,MH)P(\eta_j| \theta,\eta^*,MH) d\eta_j\,,\]
in order that one can calculate the overall likelihood $P(x_1,x_2|\theta,\eta^*,NH)=\Pi_{j=1}^2 P(x_j|\theta,\eta^*,NH)$ for further inferences. 
   

  \item We have listed \black{a few different metrics} to represent
  sensitivity of future experiments in Sec.~\ref{sec:bayes}. Each
  of them convey different information.  In the case that one has to
  choose a single number to summarize the experiment sensitivity, \black{one convenient
  choice would be $\overline{P}\equiv \overline{P}_{T=NH}^{NH}=\overline{P}_{T=IH}^{IH}$, the average probability reported for the true underlying model}. For all other metrics that were introduced, 
the sensitivity scores need to be calculated separately assuming NH or IH
is the true model.  
    
    \item For general models where
  nuisance parameters are present, it is possible to measure the
  specificity of an experiment conditional on different possible
  values of the nuisance parameters. For instance, suppose NH, and
  that a particular value of the nuisance parameter, say
  $\eta=\eta_0$, is true. Then the relevant population of potential
  experimental results consists of $x$ generated from
  $P(x|NH,\eta_0)=\int
  P(x|\theta,\eta_0,NH)P(\theta,\eta_0|NH)d\theta$. Accordingly,
  $P(NH|x)$ can be obtained for each $x$ in this population 
  with Eq.~\ref{eq:b21}~\footnote{One should not take into account
  the information of $\eta=\eta_0$ in calculating the probability, 
  since one does not know the true value of $\eta$ when analyzing
  experimental data.}, 
  and for e.g., their mean $\overline{P}_{T=NH}^{NH}(\eta_0)$ and quantiles
  $P_{T=NH}^{A}(\eta_0)$ serve as more refined sensitivity
  metrics for the experiment, and can be plotted against a range of
  possible $\eta_0$ values.
Such application is particularly 
useful when the separation of MH strongly depends on the value of $\eta$. One 
such example is long baseline $\nu_{e}$ or $\bar{\nu}_e$ appearance measurements 
(from $\nu_{\mu}$ or $\bar{\nu}_{\mu}$ beam), 
in which the sensitivity of MH strongly depends on the value of 
CP phase of lepton section $\delta_{CP}$ and neutrino mixing angle 
$\theta_{23}$.  

\item 
The Gaussian approximation in Eq.~\ref{eq:chi-bar-rms1} allows analytical calculation of various sensitivity metrics. Be aware that such calculations are valid under the assumption that the possible range of $\theta$ under either hypothesis is narrow enough that it can be reasonably represented by a single point, and that 
$\mu_i^{NH} - \mu_i^{IH} << \mu_i^{NH} \sim
\mu_i^{IH}$. 
For more general cases, numerical such as MC methods are needed. 

\item Finally, we emphasize that sensitivity metrics are designed to evaluate an experiment in its planning stage. It can be used to see if an experiment with a proposed sample size, i.e., the expected bin counts $\{\mu_i, i=1,\cdots, n\}$, will be large enough to have a high probability of generating desired strength of evidence to support the true hypothesis. But once the data are observed, the calculation of sensitivity metrics is no longer relevant. One should clearly differentiate results 
deduced from data from that from the sensitivity calculations. 
\end{enumerate}

\section{Summary}\label{sec:summary}
In this paper, we perform a statistical analysis for the problem of determining the neutrino mass hierarchy. A classical method of presenting experimental results is examined. Such method produces confidence intervals through the parameter estimation of $\Delta m^2_{32}$ based on approximating the distribution of $\sqrt{\Delta \chi^2}$ as the standard Gaussian distribution. However, due to strong
existing experimental constraints of $M^2_{32}\equiv|\Delta m^2_{32}|$, the parent distribution of the best estimation 
of $\Delta m^2_{32}$ is better approximated as a Bernoulli 
distribution rather than a Gaussian distribution, which leads to a 
very different estimation 
of the confidence level. The importance of using 
the Feldman-Cousins approach to determine the confidence interval is emphasized. 

In addition, the classical method is shown to be inadequate to convey 
the message of how much results from an experiment favor one hypothesis than the other, as the agreement between the confidence interval and the Bayesian credible interval also breaks down due to the constraints on $M^2_{32}$.

We therefore introduce the Bayesian approach to quantify the probability of MH. 
We further extend the discussion to quantify experimental sensitivities of future measurements.

\begin{acknowledgments}
We would like to thank Petr Vogel, Haiyan Gao, Jianguo Liu, Alan Gelfand, 
and Laurence Littenberg for fruitful discussions and careful reading. 
This work was supported in part by Caltech, the National Science Foundation,
and the Department of Energy under contracts DE-AC05-06OR23177, under which 
Jefferson Science Associates, LLC,  operates the Thomas Jefferson
National Accelerator Facility, and DE-AC02-98CH10886.
\end{acknowledgments}

\appendix

\section{Derivation of $P(\Delta \chi^2_{min})$ for case II:
 $\Theta=\{-1, 1\}$}\label{sec:ap1}
Let $\theta_0$ denote the true parameter value from which the data are
generated.  Under case II, when $\theta_0=1$, the statistic
$\Delta\chi^{2}_{min}(\theta_0)$ in Eq.~\ref{eq:chisqmin} is
directly related to $\Delta \chi^2$ in
Eq.~\ref{eq:delta-chi-Gaussian} (recall that the notation $\theta=1,
-1$ refers to NH and IH, respectively) as $\Delta \chi^2_{min}(1) =
\max\{0, -\Delta \chi^2\}$.  The result in section~\ref{sec:example}
implies that, under $\theta_0=1$, $-\Delta \chi^2$ follows an
approximately Gaussian distribution with mean $-\overline{\Delta
  \chi^2}$ and standard deviation $2\sqrt{\overline{\Delta
    \chi^2}}$. Similarly, when $\theta_0=-1$, the statistic
$\Delta\chi^{2}_{min}(\theta_0)= \max\{0, \Delta \chi^2\}$, where
$\Delta \chi^2$ follows approximately Gaussian distribution with mean
$\overline{\Delta \chi^2}$ and standard deviation
$2\sqrt{\overline{\Delta \chi^2}}$. Therefore, whether the truth is
$\theta_0$ is $1$ or $-1$, the distribution of
$\Delta\chi^{2}_{min}(\theta_0)$ is such that
$P(\Delta \chi^2_{min}(\theta_0)\geq t)=1$ for $t\leq 0$, and that
$P(\Delta \chi^2_{min}(\theta_0)\geq t) \approx
\frac{1}{2}-\frac{1}{2}\text{erf}\left(\frac{t+\overline{\Delta
      \chi^2}}{\sqrt{8\overline{\Delta \chi^2}}}\right)$ for $t > 0$.

\section{Confidence Interval vs. Bayesian Credible Interval}\label{sec:ap2}

As emphasized in Ref.~\cite{FC_stat}, the classical confidence
interval should not be confused with the Bayesian credible
interval. However, it is rather common that physicists approximate the
confidence interval as the Bayesian credible interval,
especially in MC simulations, where previous measurements of some physics quantities are used as inputs. Such approximations turn out to be
acceptable under the following condition.
 
Consider the condition that the pdf (or pmf) of the best estimation of
the unknown parameter $\theta_{min}$ only depends on its relative
location with respect to the true parameter value, that is,
\begin{equation}\label{eq:limit}
P_{\Tm|\Tt}(\y|\x) = h(\y-\x),
\end{equation}
for some non-negative function $h$ such that $\int_{-\infty}^\infty
h(t) dt =1$. Models that satisfy Eq.~\ref{eq:limit} are said to belong to a location family, where $\x$ is called the location parameter. When there is a lack of strong prior information for $\x$, it is usually reasonable to assign a uniform prior for it, 
that is, to assign $P_\Tt(\theta_\text{true})\propto 1$. If so, we have
\begin{widetext}
\[
\begin{split}
&P_{\Tt|\Tm}(\theta_\text{true}|\theta_{\min})\\
&=  P_{\Tm|\Tt}(\y|\theta_\text{true})P_\Tt(\theta_\text{true})/P_\Tm(\y)\\
& \propto  P_{\Tm|\Tt}(\y|\theta_\text{true})P_\Tt(\theta_\text{true})\;\;\;\text{(as a function of $\theta$)}\\
&\propto P_{\Tm|\Tt}(\y|\theta_\text{true})=h(\y-\theta_\text{true})\,.
\end{split}
\]
\end{widetext}
In the above, the first step follows from the Bayes' theorem, and the third step incorporates the uniform prior on  $\theta_\text{true}$. 
Since for any fixed $\x$, $\int_{-\infty}^\infty h(\y-\x)d\y=1$, 
the above indeed implies that \begin{equation}\label{eq:tt}
P_{\Tt|\Tm}(\y|\x)=h(\y-\x).
\end{equation}

For any threshold level $c$ and the observed value of $\theta_{\min}$, define a plausible region for $\theta_\text{true}$ by $A(\theta_{\min},c)=\{\theta:P_{\Tm|\Tt}(\theta_{\min}|\theta)>c\}$, then
\begin{equation}\label{eq:A}\begin{split}
&A(\theta_{\min},c)=\{\theta:h(\theta_{\min}-\theta)>c\}\\
=&\{\theta_{\min}+t: h(0-t)>c\}=\theta_{\min}+A(0,c)\,,
\end{split}
\end{equation}
where the transformation $t=\theta-\theta_{\min}$ is used in step~2, and 
in general, the notation $\alpha+A$ for a point $\alpha$ and a set $A$ represents the set that consists of points $\alpha+a$ for all $a\in A$.
In words, Eq.~\ref{eq:A} says that the plausible regions based on different $\theta_{\min}
$ with a fixed threshold $c$ are simply shifts in location of each other.
First, under the Bayes framework, $A(\theta_{\min},c)$ can be considered as a credible region (most often an interval). The probability that $\theta$ falls in $A(\theta_{\min},c)$ is called the level of the credible region, and is given by
\begin{widetext}
\[\begin{split}
P_{\Tt|\Tm}(\theta\in A(\theta_{\min},c)|\theta_{\min})&=\int_{A(\theta_{\min},c)} P_{\Tt|\Tm}(\theta|\theta_{\min})d\theta\\
(\text{by Eq.~\ref{eq:tt} and Eq.~\ref{eq:A}})\;\;\;&=\int_{\theta_{\min}+A(0,c)} h(\theta-\theta_{\min})d\theta\\
(\text{letting $t=\theta-\theta_{\min}$} )\;\;\; &=\int_{A(0,c)} h(t)dt\,.
\end{split}
\]
\end{widetext}
On the other hand, under the classical framework, $A(\theta_{\min},c)$ serves as a confidence interval, the level of which is given by
\begin{widetext}
\[\begin{split}
&P_{\Tm|\Tt}(\theta\in A(\theta_{\min},c)|\theta)\\
(\text{by  Eq.~\ref{eq:A}})\;\;\;=&P_{\Tm|\Tt}(\theta\in \theta_{\min}+A(0,c)|\theta)\\
=&P_{\Tm|\Tt}(\theta_{\min}\in \theta-A(0,c)|\theta)\\
(\text{by Eq.~\ref{eq:limit}})\;\;\;=&\int_{\theta-A(0,c)} h(\theta_{\min}-\theta)d\theta_{\min}\\
(\text{letting $t=\theta_{\min}-\theta$} )\;\;\; =&\int_{A(0,c)} h(t)dt\,.
\end{split}
\]
\end{widetext}
In summary, 
the region $A(\theta_{\min},c)$ can be interpreted both as a confidence interval and a credible region of the same level.

A most useful special case where Eq.~\ref{eq:limit} is satisfied is the case where $\theta_{min}$ strictly follows a Gaussian distribution with mean $\theta_{true}$ (such as Case I of section~\ref{sec:parameter}) and that the
standard deviation of the Gaussian distribution did not depend on
$\theta_{true}$.  As we mentioned in section~\ref{sec:parameter}, it is shown by Wilks~\cite{Wilks-1938} that, based on a large data sample size, the statistic $\theta_{min}$
does approximately follow a Gaussian distribution with mean at
$\theta_{true}$ under certain regular conditions. Hence, it is not unacceptable to construct an $\alpha$ level confidence interval and interpret it as an $\alpha$ level credible interval, as long as the standard deviation of the Gaussian distribution has weak or no dependence on $\theta_{true}$.

However, in the MH determination problem, the regularity conditions are violated due to the existing experimental constraints on
$|\theta|=M^2_{32}$. As a result, condition Eq.~\ref{eq:limit} is far from being satisfied, and there is no longer a correspondence between confidence intervals and Bayesian credible intervals. 
Indeed, strong inconsistency between implications of the two types of intervals can be seen from the following specific example belonging to case II of section~\ref{sec:parameter}. 
It is easy to come up with an observed data $x$ that results in $\Delta \chi^2=1$  and
$\overline{\Delta \chi^2} = 9$ (defined in Eq.~\ref{eq:chi2} and \ref{eq:chi-bar-rms1} respectively). Then, according to the Bayesian approach, the
probability is about 62.2\% that NH is the correct hypothesis,
or an odds of $5:3$ of NH against IH. Most people would consider this a fairly weak preference for NH. On the other hand, the classical estimation procedure turns out to exclude the point IH from the 95\% confidence interval according to (the correct table) Table.~\ref{table:par}. Had one attempted to interpret this 95\% confidence interval as a Bayesian credible interval, one would conclude that the odds of NH against IH is at least $19:1$. This conclusion is over confident in the MH determination
compared to the odds of $5:3$ suggested by the well-founded Bayesian approach.


\bibliographystyle{unsrt}
\bibliography{StatMH}{}

\end{document}